# Colossal electroresistance and electronic phase separation in mixed-valent manganites.

T. Wu[†], S. B. Ogale[*], J. E. Garrison, B. Nagaraj, Z. Chen, R. L. Greene, R. Ramesh and T. Venkatesan

Center for Superconductivity Research, Department of Physics, University of Maryland, College Park, MD 20742-4111.

and

A. J. Millis

Center for Materials Theory, Department of Physics and Astronomy, Rutgers University, 136 Frelinghuysen Rd, New Brunswick, NJ 08854.

Abstract :

The sensitivity of transport in CMR manganite to external fields is examined using a PZT-based field effect configuration with a $La_{0.7}Ca_{0.3}MnO_3$ channel subjected to electric and magnetic fields, separately and in conjugation. We not only find a colossal electroresistance (CER) of ~ 76 % at $4 \times 10^5$ V/cm, but also a remarkable complimentarity of this CER with the widely studied CMR. Specifically, the $ER_{max}$ ($MR_{max}$) of the system is largely independent of the magnetic (electric) field biasing, and for equal conductivity values, the ER (MR) is lower (higher) for $T<T_p$ than that for $T>T_p$. The size and systematics of the effect strongly favor a percolative phase separation picture.

Colossal magneto-resistance (CMR) – a huge decrease in resistivity by the application of magnetic field observed in mixed-valent manganites – has triggered intense scientific activity in recent years; yet, the mechanism of the effect is still not fully understood[1-3]. One aspect of the phenomenon is a bulk metal to insulator transition tuned by changing spin correlations, but possibly involving other degrees of freedom also.[4] Recent studies however suggest that the largest magnetoresistance in these systems is associated with spatial inhomogeneity related to multiphase coexistence.[5-6] Multiphase coexistence generically causes a sensitivity of physical properties to external perturbations. We examine the sensitivity to external fields in the case of CMR manganite using a PZT-based field effect configuration with a $La_{0.7}Ca_{0.3}MnO_3$ channel subjected to electric and magnetic fields, separately and in conjugation.

An early study of an electric field effect using a dielectric gate revealed interesting electroresistance phenomena[7], and this was followed by the demonstration of a large field effect in a manganite channel using a ferroelectric gate[8,9]. In neither of these studies, however, was the temperature or the magnetic field dependence of the phenomenon examined. There have been some interesting reports on current induced switching of resistive states in manganites[10-13], also referred to as field effects. However, these studies were on charge-ordered samples, and involved injection of current into apparently highly filamentary current paths due to intrinsic or induced inhomogeneities. The possibility of substantial thermal effects on a microscopic scale renders interpretation difficult in such situations. In this work we have fabricated a significantly improved, inverted PZT-based CMR-FET device, with which we have discovered remarkable new features pertaining to the relationship between the electro- and magneto-resistance in the system. Our approach is non-invasive: no current is injected, and there is no concern of local heating as being a possible cause of any of our observations. We not only find a colossalelectroresistance (CER) of ~ 76

% at 4 × $10^5$ V/cm, but also a remarkable complimentarity of this CER with the widely studied CMR. The size and systematics of the effect strongly favor a percolative phase separation picture.

In the inset of Fig. 1 we show our device configuration. A (100)-orientated single crystal of n-type STO doped with 1 at. % Nb serves as the substrate for the growth of the heterostructure as well as the conducting gate. The surface preparation procedure outlined by Kawasaki et al.[14] was followed, which was found to be crucial for reliably obtaining high quality devices with significantly improved performance. Both the 150nm layer of PZT and 50 nm layer of LCMO were grown by pulsed laser deposition at 650°C, under $O_2$ pressure of 100 mTorr and 400 mTorr, respectively. In this inverted configuration the channel layer is deposited on the gate dielectric layer, which is deposited on the conducting substrate that serves as the gate. This differs from the configuration used in the previous studies[7-9] wherein the active channel layer was deposited first on an insulating substrate, followed by the deposition of the dielectric and metal contacts. The presently used inverted configuration offers several advantages, the most important being the significant reduction in the processing steps leading to improved device performance and high reproducibility.

Fig. 1 shows the dependence of resistivity of the CMR channel on temperature, for field biasing over an applied gate voltage range from +6 to –6 volts. Not only is the effect much stronger ( $ER_{max}$ = [R(E)–R(0)]/R(0) = 76 % for E = 4 × $10^5$ V/cm near $T_p$) than reported in the previous studies[7-9], but these data also reveal entirely new features. Firstly, the change in resistance changes sign with alteration of the field direction. Secondly, the strength of the effect is not symmetric vis a vis the field direction, which implies that the channel exhibits asymmetry for hole vs electron type carrier modulation. Thirdly, there is hardly any shift in the peak temperature $T_p$ up to 6V, which is in sharp contrast with the known shift towards higher temperature upon the application of a magnetic field. Also, the electric field effect is notably large even above $T_p$. Clearly, the character of E-field influence on the CMR state is distinctly different as compared to that due to H-field. We note that we also found similar features in devices with a linear dielectric gate, namely $SrTiO_3$ as shown in Fig. 2. This experiment allowed us to examine the channel resistance under a more linear dielectric response, unlike the case for PZT where the polarization states are discrete. As shown in the inset, the channel resistance increased and saturated for positive bias, but for the negative bias a nonlinear decrease with no evidence for saturation was seen, consistent with the PZT data.

In Fig. 3 we present data on the combined consequences of the electric and magnetic fields for the transport in the CMR channel. Specifically, we show $\rho$ vs T for the unbiased and field-biased ($4\times10^5$ V/cm) channel, with and without 6 T field. We observe that an electric field of $4\times10^5$ V/cm applied to the channel produces a change in resistivity which is almost as large as that produced by a magnetic field of 6 T. Thus, the electroresistance (E-field effect) in this system is as colossal as the magnetoresistance. Further, it is seen that the application of a magnetic field of 6 T to the E-field-biased channel produces an additional large drop in resistivity (i.e. change from curve B to D). Similarly, application of electric field to the channel subjected to a magnetic field of 6 T also produces an additional large drop in resistivity (i.e. change from curve C to D). In both cases, application of magnetic field shifts the peak of $\rho$, and of the ER. We show in the insets MR Vs T in 6 T for the unbiased and E-field-biased CMR channel, and ER vs T for the channel with and without a magnetic field of 6 T. The difference seen in the ER case arises from the peak shift characteristic of MR. It is extremely interesting and remarkable that the magnitudes of MR and ER are almost comparable for the two conditions compared in each figure. Thus, the electric and magnetic fields have a complimentary effect on the film transport. This is borne out even more strongly by the dependence of MR and ER on $\sigma$ as shown in Fig. 4. In plotting these data the conductivity, and the MR and ER values at each temperature are used. In the inset of fig. 4 (a) the behavior of ER near the metal-insulator transition is shown enlarged. For equal $\sigma$ values, the ER is lower for $T<T_p$ than that for $T>T_p$. Interestingly, this behavior is exactly opposite for MR, which once again emphasizes the complimentary nature of the influences of E and H fields.

We now consider the interpretation of our results. That the "gate" electric field has any effect at all on the resistivity is already remarkable. If the carrier density were equal to the chemical density (0.7 electrons/Mn $\cong 10^{22}cm^{-3}$) any applied electric field would be screened within one or two lattice constants of the interface, and could not affect bulk transport. In our films, therefore, some process has opened a gap in the electronic spectrum, thereby reducing the effective carrier density by orders of magnitude and allowing the electric field to penetrate deeply into the film. Of course this gap is also evident in the rapid temperature dependence of resistivity. The gap could be due to polaron formation, i.e. electron trapping by spatially incoherent lattice distortions, as proposed by Millis et al.[15] and Alexandrov et al.[16], but in these models it is difficult to obtain the very rapid upturn of $\rho$ between 300 and 200 K, and the very long screening lengths we observe.[17] We believe our data are more naturally interpreted as a consequence of charge ordering in our films. While charge ordering has not been observed in bulk samples of the composition we studied, it can be induced by modest chemical doping, for example substitution of La by Pr[5] and it has been suggested that strain, microstructure and ion damage can also induce an inhomogeneous charge ordered state[18], similar to the 'multiphase' state found in (La/Pr)$CaMnO_3$.[5]

The applied field converts some portion of the film into a metallic state, presumably by rearranging the charge in some way. The usual electroneutrality requirements mean that positive gate voltages cause electrons to accumulate at the gate electrode and holes to accumulate at the top of the film, and the change in the charge density causes one or the other side to become metallic. If the system were homogeneous, the created metallic region would be very thin because it would again screen any induced electric field on a very short length scale. We believe that it is much more likely that the film (at V=0) is in an inhomogeneous state, partly metallic and partly insulating. In order to obtain the inhomogeneity, it is necessary that the insulating behavior be due to charge/orbital ordering. At V=0 and $T>T_p$ the metallic portions do not percolate. A gate voltage then causes charge to accumulate between metallic and insulating phases, and this change causes the interface to move, increasing the volume fraction of one phase at the expense of the other.

We now make this picture more precise. We assume that the V=0 material is similar to $(La_{1-x}Pr_x)_{0.7}Ca_{0.3}MnO_3$,[5] which indeed has a ρ(T) similar to that of the film we studied. This material exhibits multiphase behavior, with charge order and ferromagnetic metal coexisting. In this system, x<0.3 is more metallic than x>0.3, because the charge order is of the 0.5:0.5 type. Thus, an accumulation of holes at an interface between the metallic and insulating regions will move that interface deeper into the metallic phase, increasing the volume of the insulator. A negative gate voltage causes accumulation of electrons at the interface whereby the interface is moved into the insulating phase resulting in an increased metallicity of that region. This increases the connectivity and hence the conductivity. In the percolation regime such increase should have a non-linear dependence on the gate voltage, as observed. The effect would only saturate when all the insulating regions are driven out of the top part of the film. On the other hand, a positive gate voltage moves electrons towards the bottom of the film, making the region near the source and drain electrodes more insulating.

Clearly, this effect should saturate once the connectivity of the percolation system is mostly broken, as also observed.

Within this picture the complimentary character of the E and H field influence on the state of the CMR channel is naturally explained. The H-field aligns the relative orientations of magnetizations of the FM regions with reference to each other giving the CMR effect, while the E-field changes the connectivity of the metallic regions by modifying the volume fractions of the FM and CO components. Since the two fields operate entirely differently, the maximum MR is expected to be largely independent of the presence or absence of E-field and the maximum ER is also expected to be mostly independent of the presence or absence of the H-field, as is indeed observed.

In the argument above, the electric field is suggested to penetrate the 50:50 type charge ordered regions in the film and move the interface between such regions and the ferromagnetic ones, rather than changing the character of either phase. In order to rule out the possibility of any direct modification of the transport in the charge ordered regions themselves, we also performed the field-effect experiment on a $La_{0.5}Ca_{0.5}MnO_3$ film. The corresponding data are shown in Fig. 5. Clearly there is a very small E-field effect on the bulk transport in this channel for comparable field, over the entire temperature range. We also note that in addition to applying field through a gate as done for the data shown in Fig. 5, we instead injected current in the channel itself as done byAsamitsu et al.[10] and others[11-13]. In this case we obtained changes similar to those seen by these authors, which we believe are heating effects due to filamentary current paths. This establishes the non-invasive nature of our gate approach and its potential in bringing out delicate non-thermal aspects of the problem.

The notion of an electronically phase separated channel is entirely new to the area of field effect studies and could potentially lead to novel device concepts. Studies of composite systems involving internal modulation of electronic properties have revealed that the electric fields in such cases are extremely inhomogeneous[19-21] and can lead to a large enhancement in the effective linear and non-linear responses[20]. Indeed a number of surface enhanced optical non-linearities responsible for such phenomena as the Kerr-effect, four wave mixing and harmonic generation have been demonstrated in the case of metal-dielectric composites[21].

In conclusion, we have observed a CER in a CMR channel subjected to electric field through a PZT gate, exhibiting peculiar characteristics in the presence and absence of an applied H-field. All our observations find a consistent explanation in the two-phase picture of transport in manganites. In addition to its obvious relevance to the fundamental aspects of the mechanism of the CMR effect, our study should also stimulate a development of novel FET device concepts based on the notion of using intrinsically or artificially inhomogeneous channels.

This work was supported by NSF-MRSEC under grant number DMR-96-32512 and DMR 9705482 (AJM).

References :

† Electronic mail : tomwu@squid.umd.edu

* Electronic mail : ogale@squid.umd.edu

Figure Captions :

Fig. 1 : Dependence of resistivity of the CMR channel on temperature, for field biasing with a PZT gate over an applied gate voltage range from +6 to –6 volts. Inset shows the device configuration.

Fig. 2 : Dependence of resistivity of the CMR channel on temperature, for field biasing with an STO gate over an applied gate voltage range from +3 to –3 volts. Inset shows the channel resistance dependence on gate voltage.

Fig. 3 : Dependence of channel resistivity on temperature for the unbiased (A) and electric field-biased (B, $4\times10^5$ V/cm) channel, in the absence of magnetic field. The dependence (A) changes to (C), and (B) changes to (D) under a magnetic field of 6 T. The insets show MR Vs T in 6 T for the unbiased and E-field-biased CMR channel, and ER Vs T for the channel with and without a magnetic field of 6 T.

Fig. 4 : Dependence of a) electro-resistance (ER) and b) magneto-resistance (MR) on the conductivity of the CMR channel. Inset of a) shows the behavior of ER near the metal-insulator transition on expanded scale.

Fig. 5 : Dependence of resistivity of the $La_{0.5}Ca_{0.5}MnO_3$ channel on temperature, for field biasing with a PZT gate over an applied gate voltage range from +6 to –6 volts.

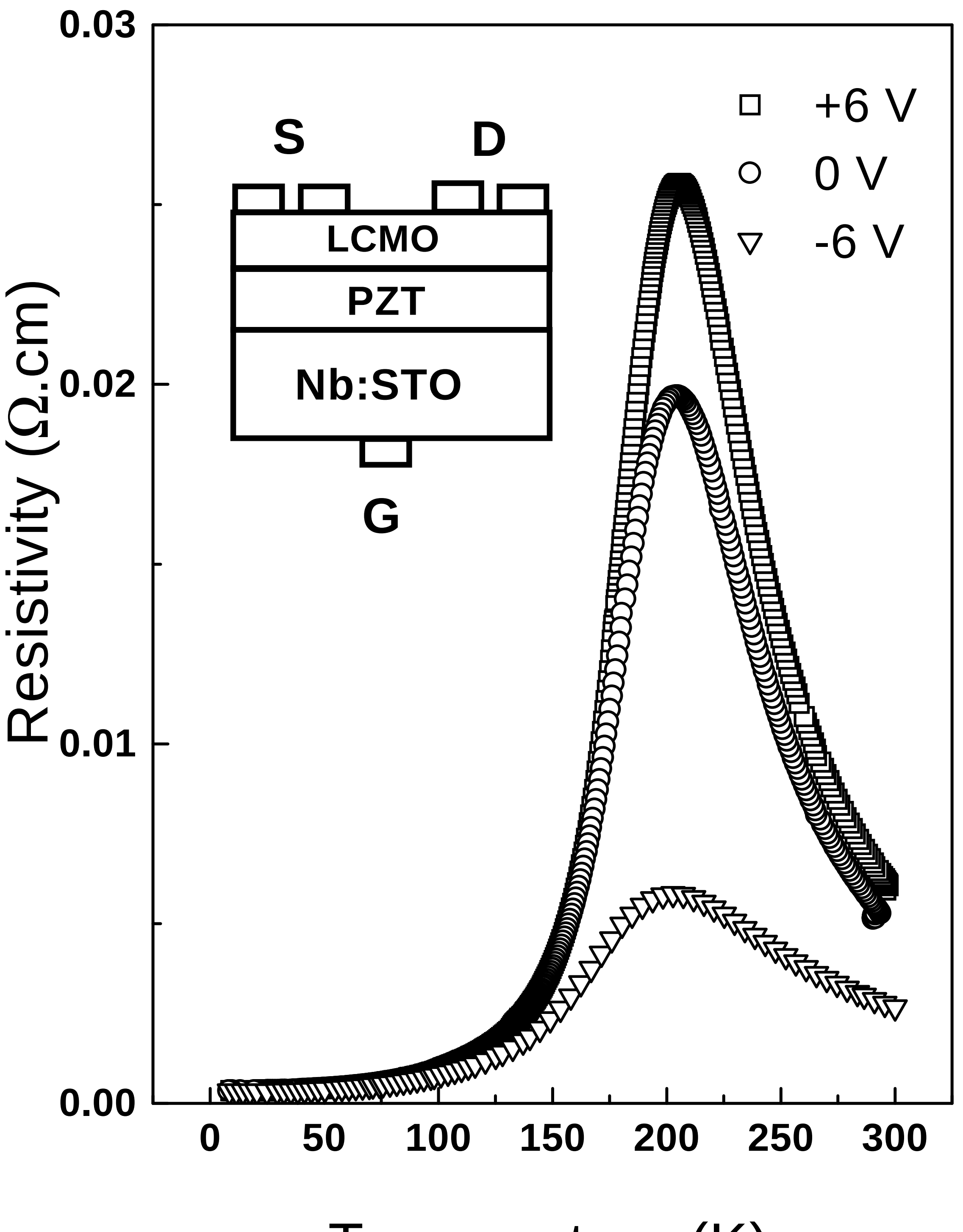

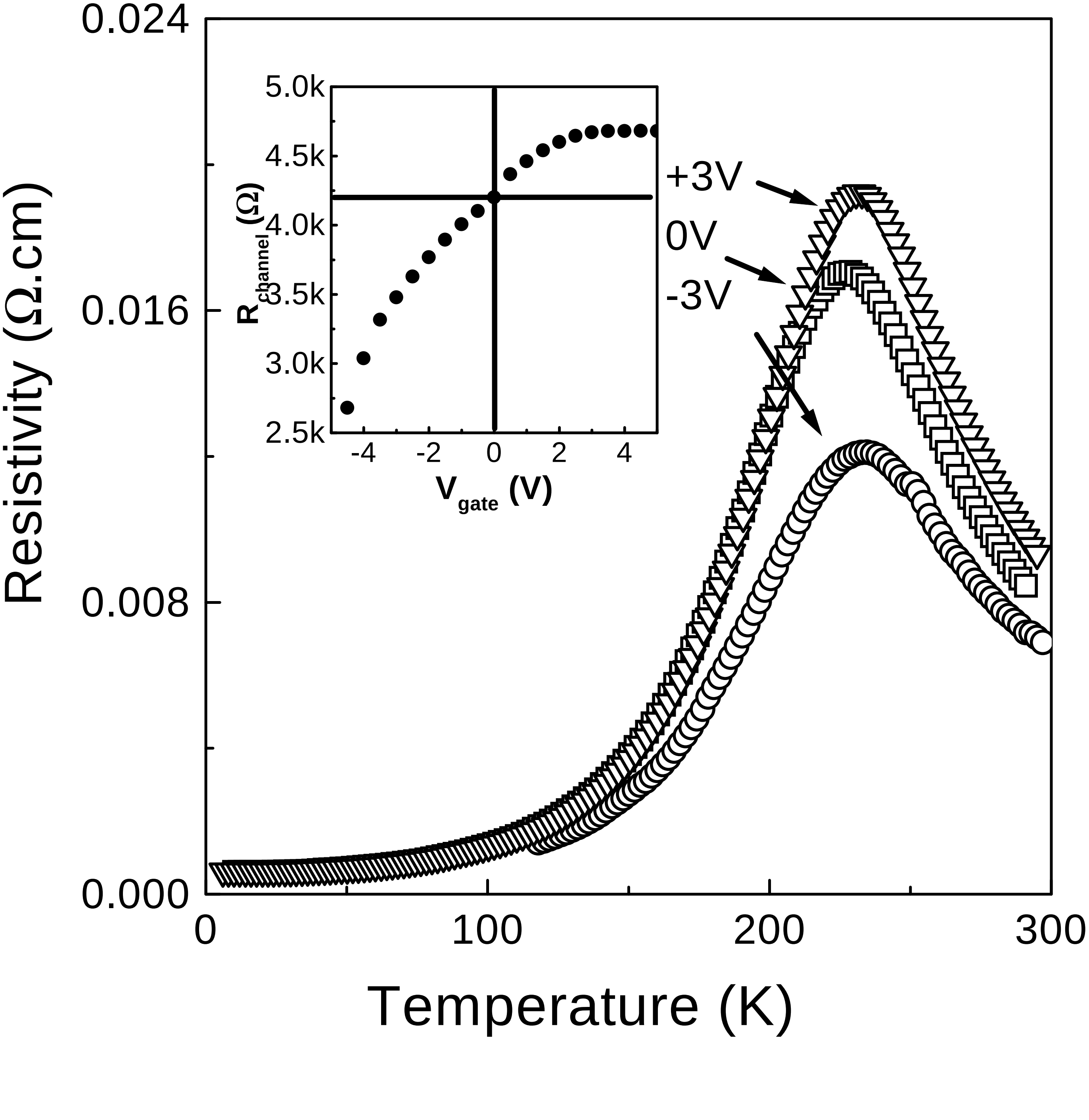

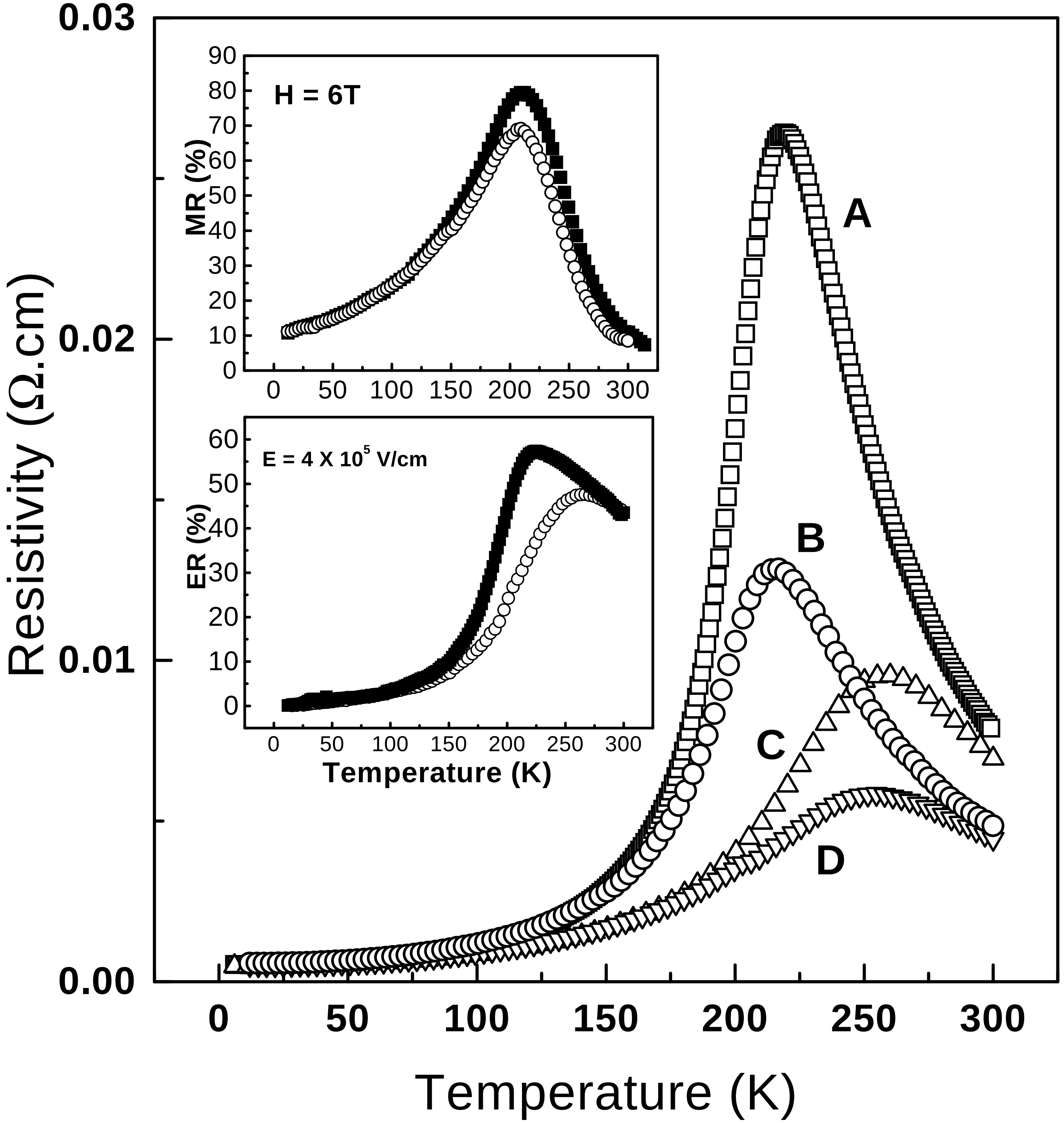

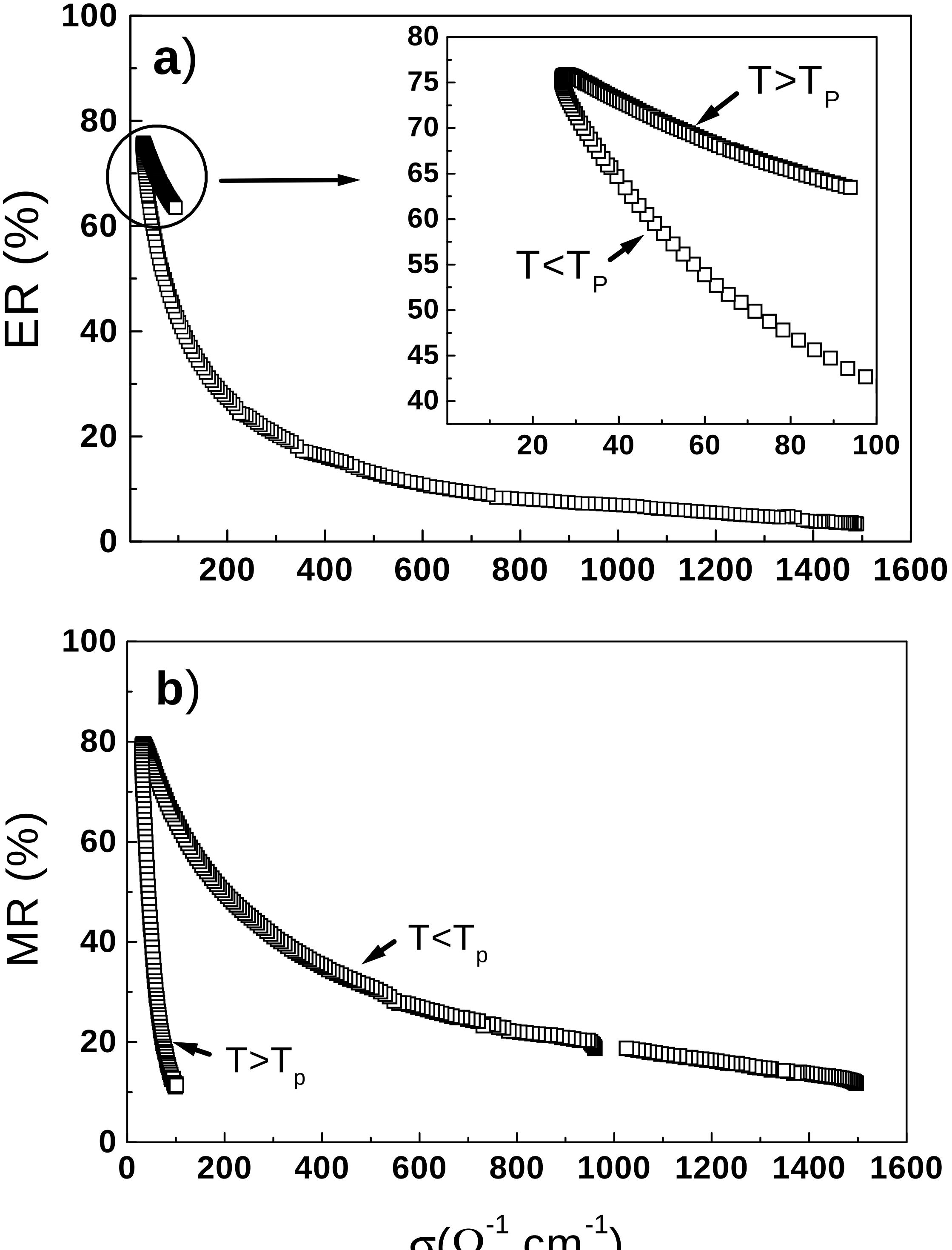

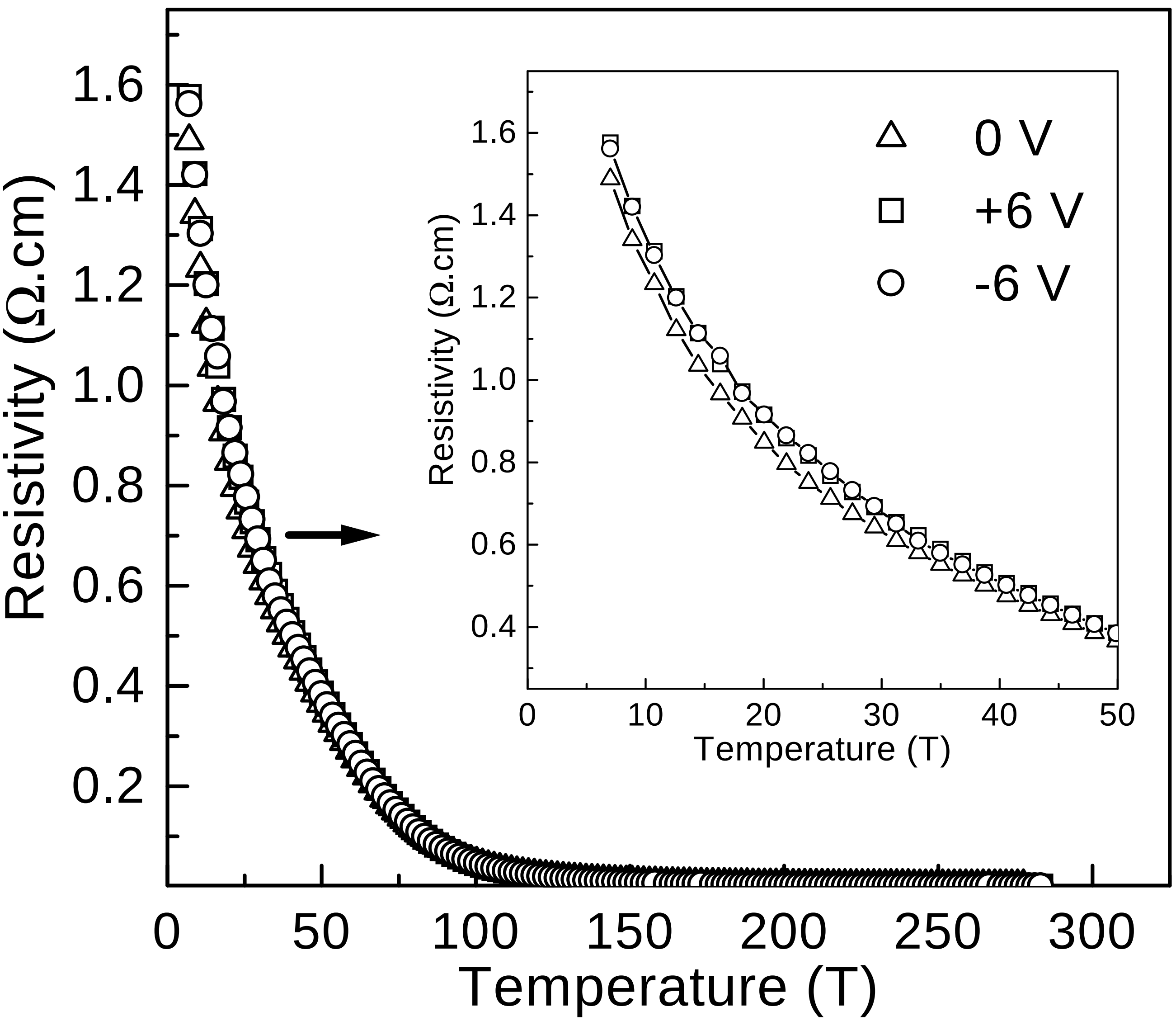